# Internal Modes of Oscillations of Topological Solitons in Highly Dispersive Media

## O.V. Charkina, M.M. Bogdan

*B. Verkin Institute for Low Temperature Physics and Engineering of the National Academy of Sciences of Ukraine, 47 Lenin Ave., Kharkov, 61103, Ukraine*

E-mail: charkina@ilt.kharkov.ua , bogdan@ilt.kharkov.ua

The problem of stability and spectrum of linear excitations of a soliton (kink) of the dispersive sine-Gordon and $\varphi^4$ - equations is solved exactly. It is shown that the total spectrum consists of a discrete set of frequencies of internal modes and a single band spectrum of continuum waves. It is indicated by numerical simulations that a translation motion of a single soliton in the highly dispersive systems is accompanied by the arising of its internal dynamics and, in some cases, creation of breathers, and always by generation of the backward radiation. It is shown numerically that a fast motion of two topological solitons leads to a formation of the bound soliton complex in the dispersive sine-Gordon system.

## Introduction

The soliton theory of nonlinear systems in the long wave limit is developed quite well [1,2]. Typical examples of soliton excitations are dislocations in an imperfect lattice [3], fluxons (magnetic fluxes) in arrays of the Josephson junctions [4], magnetic and ferroelectric domain walls, and nonlinear excitations of microtubules in nerve tissue cells [5]. These topological solitons are described in the long-wavelength limit in the framework of the usual sine-Gordon and $\varphi^4$ - models [1,2]. Discrete versions of the systems demonstrate new properties, which are known as the discreteness effects [6,7]. Such effects in the discrete sine-Gordon system are described by the equation [7]:

$$\frac{\partial^2 u_n}{\partial \tau^2} + 2u_n - u_{n-1} - u_{n+1} + \frac{1}{d^2}\sin u_n = 0, \qquad (1)$$

where $d$ is a discreteness parameter. Since the frequency of linear waves $\omega$ depends strongly on the wave number $k$:

$$\omega(k) = \sqrt{d^{-2} + 2(1-\cos k)}, \qquad (2)$$

the discrete models are naturally related to a class of highly dispersive systems. The strong spatial dispersion in lattice systems can change drastically the dynamic properties of nonlinear topological excitations. In the medium with strong dispersion the nonlinear waves exhibit a complex intrinsic structure, which manifests itself in their ability to be flexible. As a result, discrete solitons can possess internal degrees of freedom and hence internal modes of oscillations [8]. Another example of highly dispersive models is a continuous sine-Gordon model with non-local but short range interaction [9] which is often employed for description of dynamical behavior of excitations in the long Josephson junctions.

Analytically the strong dispersion can be taken into account by adding the fourth-order spatial derivative in the expansion of the second difference

$$u_{n-1} + u_{n+1} - 2u_n \approx u_{xx} + \gamma u_{xxxx}, \qquad (3)$$

where $x = n/d$ and the parameter $\gamma = 1/12d^2$. However the use of the equations with the fourth-order spatial derivative leads to an artificial instability of the $u = 0$ state with respect to short waves generation. In fact, the dispersion relation for linear waves in the equation has the form:

$$\omega(q) = \sqrt{(1 + q^2 - \gamma q^4)}, \qquad (4)$$

and it is easy to see that the frequency becomes zero at the finite $q_0$. To avoid the instability, Boussinesq's primary idea of using the mixed spatio-temporal derivative, instead of the spatial one, has been applied to the dispersive sine-Gordon and double sine-Gordon equations [10-13]. Rosenau [14] has argued for this substitution in the lattice theory, and presently this method is actively revised [15, 16] in the theory of the discreteness effects. In particular, it has been also shown [17] that this approach is effective under description of the discrete breather effects in the Fermi-Pasta-Ulam lattice.

In this work we study topological solitons and their bound complexes in the framework of the sine-Gordon and $\varphi^4$-equations with the mixed spatio-temporal derivatives. We find exact solutions of these equations corresponding to static kinks and a moving soliton complex. We obtain explicit expressions for frequencies of internal modes of kinks of the equations and show numerically how they manifest themselves in the single kink dynamics. Then we study kink interactions in the dispersive sine-Gordon system and find the conditions of a formation of bound two-soliton complexes in the highly dispersive system.

**Dispersive nonlinear equations and exact soliton solutions**

In this section we introduce the dispersive sine-Gordon equation (dSGE):

$$u_{tt} - u_{xx} - \beta u_{xxtt} + \sin u = 0, \qquad (5)$$

and the dispersive $\varphi^4$ - equation

$$\varphi_{tt} - \varphi_{xx} - \beta \varphi_{xxtt} - \varphi + \varphi^3 = 0, \qquad (6)$$

and present their exact soliton solutions. The constant $\beta$ in the equations (5) and (6) is called the dispersive parameter. The equation (5) has the dispersion relation of the form

$$\omega_1(k) = \sqrt{(1 + k^2)/(1 + \beta k^2)} \qquad (7)$$

and hence stable states with $u = 0, 2\pi, 4\pi...$ [13] in contrast to the equation with the fourth spatial derivative $u_{xxxx}$. The equation (4) has two equivalent stable states with $u = \pm 1$ and the dispersion relation

$$\omega_2(k) = \sqrt{(2 + k^2)/(1 + \beta k^2)}, \qquad (8)$$

The first peculiarity of the both spectra is a finite frequency band for continuum waves. It should be also noted that with the increase of the parameter $\beta$ both spectra bands shrink and become more and more narrow. When $\beta = 1$ and $\beta = 1/2$ for the dSGE and the dispersive $\varphi^4$ - equation the continuum wave spectra degenerate to the frequency values $\omega = 1$ and $\omega = \sqrt{2}$, respectively. For

$\beta > 1$ and $\beta > 1/2$ the spectra have maxima at $k = 0$ and, in general, take forms of optical vibration branches which are typical for diatomic systems. In fact, it can be show that the slow-varying envelope of optical oscillations in the diatomic system obeys the equation with the fourth-order mixed spatio-temporal derivative.

The dispersion relation for the dSGE at small $k$ takes the form, which is similar to the expansion of the expression (4) for the model with the spatial derivative in the long wave limit:

$$\omega(k) \approx 1 + \frac{1}{2}(1-\beta)k^2 - \frac{1}{8}(1-\beta)(1+3\beta)k^4. \tag{9}$$

By comparison of the expressions we find the relations between parameters of the two models: $q = \sqrt{1-\beta}k$ and $\gamma = \beta/(1-\beta)$.

When $\beta = 0$, both equations become the usual Lorentz-invariant sine-Gordon and $\varphi^4$-equations. Their simplest soliton solutions, kinks, are well-known [1]. These solutions describe inhomogeneous states in which topological solitons connect two minima of the external potentials of the models. It is easy to see that static kink solutions of the usual sine-Gordon and $\varphi^4$-equations remain exact solutions of the dispersive equations (5) and (6). They have the forms

$$u_{2\pi} = 4\arctan(\exp(x)), \tag{10}$$

$$\varphi_0(x) = \tanh\left(\frac{x}{\sqrt{2}}\right), \tag{11}$$

respectively. In the case of $\beta = 0$ the moving solutions are obtained by the use of the Lorentz transformation of the coordinate $z = (x - Vt)/\sqrt{1-V^2}$ in the expressions (10) and (11). However it has been shown [12, 13] that the moving single kinks are absent as exact solutions in the case of $\beta \neq 0$. The equations for moving solutions are written as

$$u_{zz} + \alpha u_{zzzz} - \sin u = 0, \tag{12}$$

$$u_{zz} + \alpha u_{zzzz} - u + u^3 = 0, \tag{13}$$

where the parameter $\alpha$ is defined as follows

$$\alpha = \frac{\beta V^2}{(1-V^2)^2}. \tag{14}$$

This parameter plays an important role in understanding the dynamics of the dispersive models. It is clear that for small $\beta$ and $V$ the dynamical properties of the models have to be close to those of the usual Lorentz-invariant systems. The formal perturbation theory, considering $\alpha$ as a small parameter, gives e.g. for the kink of the equation (12) in the first approximation:

$$u(z) = u_{2\pi}(z) + \alpha\left(3\frac{\sinh z}{\cosh^2 z} - \frac{z}{\cosh z}\right) + \dots . \tag{15}$$

It seems that the asymptotic series could be constructed with arbitrary accuracy [18]. Nevertheless, the series (15) and a similar one for the equation (13) do not converge therefore equations (12) and (13) do not possess solutions exactly satisfying boundary conditions for single kinks. Numerical simulations confirm this fact (see below).

At the same time in the case of the dispersive sine-Gordon equation the moving bound two-soliton complex solutions can exist. One of them has been found analytically [10]

$$u_{4\pi}(x) = 8\arctan\left\{\exp\left(\sqrt{\frac{2}{3}}\frac{x-V_0 t}{\sqrt{1-V_0^2}}\right)\right\}, \qquad (16)$$

where the velocity of the complex is the function of the dispersive parameter:

$$V_0(\beta) = \sqrt{1+\frac{\beta}{3}} - \sqrt{\frac{\beta}{3}}. \qquad (17)$$

Other soliton complex profiles can be found numerically while their velocities comprise the discrete set of definite values [12, 13].

### Spectra of linear excitations of kinks in highly dispersive systems

Here we find the spectra of linear excitations of the static kinks given by the expressions (10) and (11). The problem of the spectrum for the dSGE is formulated as the following. We seek for the solution of the equation (5) in the form

$$u(x,t) = u_{2\pi}(x) + \Delta u(x,t) = u_{2\pi}(x) + f(x)\exp(i\omega t), \qquad (18)$$

where $\Delta u(x,t)$ is assumed to be small with respect to the kink solution. The linearized equation for the function $f(x)$ is the Schroedinger equation with the well-known potential well [19]

$$\left\{-(1-\beta\omega^2)\frac{d^2}{dx^2} + 1 - \frac{2}{\cosh^2 x}\right\}f(x) = \omega^2 f(x). \qquad (19)$$

One can construct a complete set of solutions of the equation (19) using results for the eigenvalue problem presented in [19]. It is evident that the continuum spectrum of the problem is given by the expression (7). The discrete values of the internal modes frequencies can be found from the equation for the discrete energy levels:

$$\sqrt{\frac{1}{4} + \frac{2}{1-\beta\omega^2}} - \sqrt{\frac{1-\omega^2}{1-\beta\omega^2}} = n + \frac{1}{2}, \qquad (20)$$

where $n$ is an integer. The number of levels is determined by the value of the parameter $\beta$. The zero eigenvalue $\omega_0 = 0$ corresponds to $n=0$ and the ground state eigenfunction describes the translational mode. With the increase of the parameter $\beta$ all the internal modes frequencies detach consequently from the low edge of the continuum spectrum. The threshold value of the parameter $\beta$ for $n$-th mode is given by the expression:

$$\beta_n = 1 - \frac{2}{n(n+1)}. \qquad (21)$$

The expressions for the frequencies of all modes can be written explicitly. For example, the first internal mode squared frequency has the form:

$$\omega_1^2 = \frac{1}{\beta}\frac{\Delta^2 - 9}{\Delta^2 - 1}, \qquad \Delta(\beta) = \frac{6\beta + \sqrt{17\beta^2 - 10\beta + 9}}{1+\beta} \qquad (22)$$

The mode detaches from the continuum spectrum at small $\beta$ as follows

$$\omega_1 \approx 1 - \frac{2}{9}\beta^2 - \frac{10}{81}\beta^3 + \dots \quad (23)$$

Its behavior coincides qualitatively with the frequency dependence on the discreteness parameter in the discrete sine-Gordon system [8, 15, 20]. It should be noted that in the case of the continuum spectrum degeneration ($\beta = 1$) the whole infinite set of the frequencies of internal modes is expressed very simply:

$$\omega_n = \sqrt{1 - \frac{2}{(n+1)(n+2)}}. \quad (24)$$

The problem of the linear excitation spectrum of the dispersive $\varphi^4$-model kink is solved similarly. At first we derive the linearized equation for the function of small deviations from the kink shape

$$\Delta\varphi(x,t) = \varphi(x,t) - \varphi_0(x) = \psi(x)\exp(i\omega t), \quad (25)$$

and, finally, the equation for $\psi(x)$:

$$\left\{-(1-\beta\omega^2)\frac{d^2}{dx^2} + 2 - \frac{3}{\cosh^2(x/\sqrt{2})}\right\}\psi(x) = \omega^2\psi(x). \quad (26)$$

The equation for the spectrum eigenvalues can be written as follows

$$\sqrt{\frac{1}{4} + \frac{6}{1-\beta\omega^2}} - \sqrt{\frac{2(2-\omega^2)}{1-\beta\omega^2}} = n + \frac{1}{2}. \quad (27)$$

The translational mode with $n=0$ and $\omega_0 = 0$ exists in general at $\beta \neq 0$. The first internal mode ($n=1$) corresponds to the oscillation of the soliton width. The expansion of the frequency of this mode at small $\beta$ has the form:

$$\omega_1(\beta) \approx \sqrt{\frac{3}{2}}\left(1 - \frac{7}{20}\beta - \frac{561}{4000}\beta^2 + \dots\right) \quad (28)$$

In the case $\beta = 0$ it is reduced to the well-known shape mode of the usual $\varphi^4$-model:

$$\Delta\varphi(x,t) = \psi_1(x)\sin(\omega_1 t) = \frac{\sinh(x/\sqrt{2})}{\cosh^2(x/\sqrt{2})}\sin\left(\sqrt{\frac{3}{2}}t\right). \quad (29)$$

The expression for the moving $\varphi^4$-kink with this internal mode can be obtained by the use of the Lorentz transformation: $z = (x - Vt)/\sqrt{1-V^2}$ and $\tau = (t - Vx)/\sqrt{1-V^2}$. It is evident that this mode does not influence on the translational motion of the mass center, and the velocity of the kink is constant.

In a dispersive system the excited kink undergoes simultaneous oscillations of its width and velocity. Even in the weak dispersion limit the kink plays the role of the moving source of radiation and can produce linear excitation with the wavelength $\lambda = 2\pi/k_0$ where the wave number is determined from equation $\omega(k_0) = Vk_0$. With the dispersion increase the effect of internal modes grows and the contribution of the continuum waves diminishes. First of all, this result can be applied to the interpretation of the kink motion in the dispersive $\varphi^4$-model even in the case $\alpha \ll 1$ due to the existence of the well-defined shape mode (29). Besides this internal oscillation, the second internal mode in the dispersive $\varphi^4$-model detaches from the continuum spectrum as follows:

$$\omega_2(\beta) \approx \sqrt{2}\left(1 - \frac{18}{25}\beta^2 - \frac{468}{625}\beta^3 + ...\right) \quad (30)$$

Thus one can expect that in the case of the small dispersion the kink motion in the dispersive sine-Gordon system can cause the excitation of one weakly-localized internal mode with frequency of Eq. (22), whereas in the dispersive $\varphi^4$-model the moving kink can be accompanied by two excited internal modes. In the strong dispersion limit the influence of the internal modes becomes crucial. The predictions of the theory are verified numerically in the next section.

### Numerical simulations of the topological soliton dynamics

We have investigated the kink dynamics in the dispersive sine-Gordon and $\varphi^4$-model by computer simulation. To model the dispersive equations, we have used a highly stable difference scheme which is similar to the one proposed in [9]. Typically we have used time step $\Delta t = 0,0001$ and spacing $\Delta x = 0,02$ in a system consisting of 3000 sites. The initial expressions for the soliton evolution have been taken from equations (10), (11), (15) and (16). Typical values of the dispersive parameter $\beta$ have been 1/12, 1/6, 1/4 and 1, and the initial velocities $V_{in}$ have been chosen in the interval between 0,3 and 0,9.

Our results are summarized as the following. At very small values of the dispersive effective parameter $\alpha \leq 0,01$ we have found that the dynamical properties of a single kink in both models differ slightly from those in the usual Lorentz-invariant equations. However, with increasing velocity the new effects take place while details of the kink motions have appeared important. First of all we have calculated the velocity of the kink mass center as a derivative of the kink center coordinate on time. The result for the dSGE kink at $\beta = 1/12$ и $V_{in} = 0,5$ is presented in Fig.1. One can see that the kink velocity begins to decay rapidly and then it oscillates, approaching to another quasi-equilibrium value. The kink tail also slightly oscillates but no forward radiation is observed. As a whole the kink evolution can be described in terms of the combined mode of translational and internal motions affected by the radiative dissipation effect. For larger initial velocity ($V_{in} = 0,6$) the kink oscillating tail becomes visible (Fig.2) even if we have started from the improved kink profile (15). The role of the radiation is much less for small $\alpha$ in the dispersive $\varphi^4$-model. For this case the velocity of the $\varphi^4$-kink center is indicated in Fig.3 when the initial value has been chosen as $V_{in} = 0,3$. Instead of the velocity decay we see the two-frequency modulated velocity. This fact reflects the existence of two internal modes (28) and (30) in the $\varphi^4$-kink spectrum and leads to generation of a complex breathing excitation.

When the parameter $\beta$ is not small and the velocity is large enough the dynamics of the kinks at the first stage becomes strongly non-stationary and dissipative as in the highly discrete systems [7]. A new channel of the kink energy loss arises due to the breather appearance at the kink wake.

This process is demonstrated in Fig.4 for the $\varphi^4$-system, where three consequent temporal positions of two excitations, the stationary kink and the static breather mode, are shown.

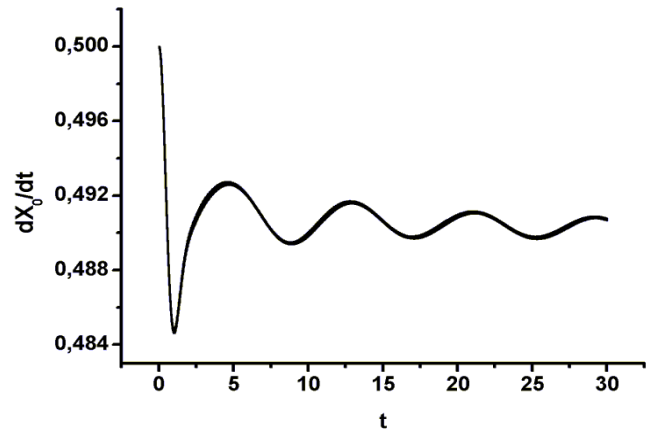

**Fig.1.** *The temporal dependence of the velocity of the kink center mass in the dispersive sine-Gordon equation at $\beta = 1/12$ and $V_{in} = 0,5$.*

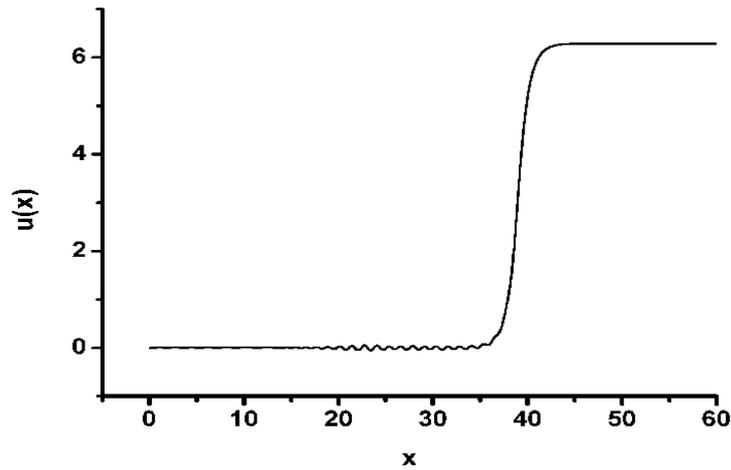

**Fig.2.** *The moving kink profile with the oscillating tail in the dispersive sine-Gordon equation at $\beta = 1/12$ and $V_{in} = 0,6$.*

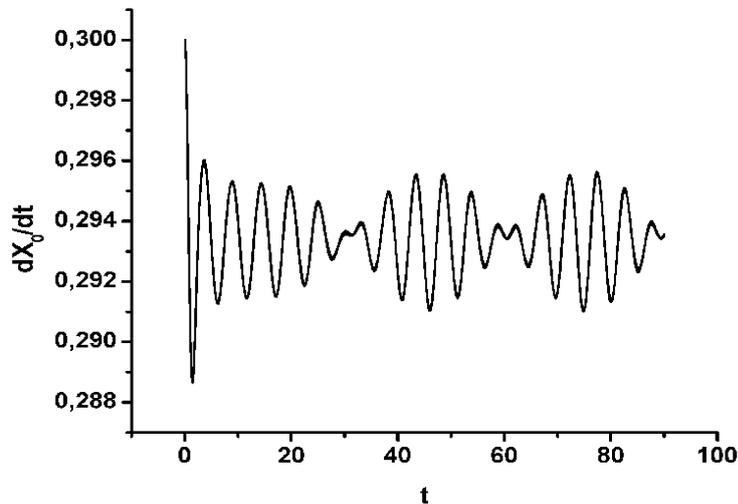

**Fig.3.** *The temporal dependence of the velocity of the kink center mass in the dispersive $\varphi^4$-equation at $\beta = 1/12$ and $V_{in} = 0{,}3$.*

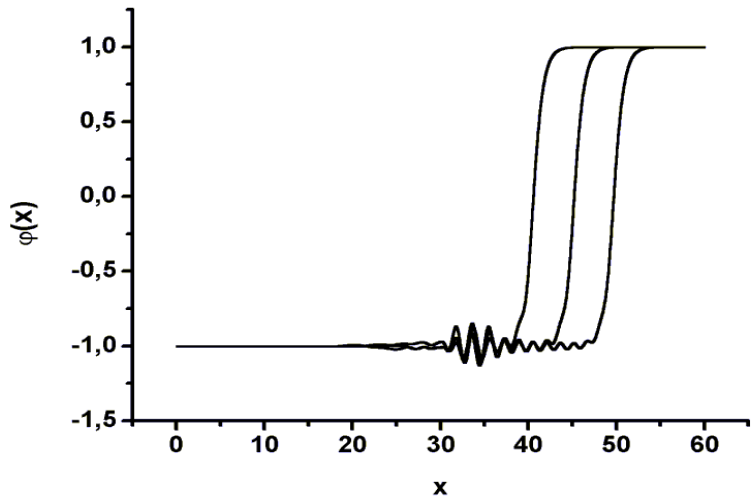

**Fig.4.** *The moving kink and the stationary breather mode in the dispersive $\varphi^4$-system.*

At the same time if the parameter $\alpha$ is large due to the large velocity, a formation of the bound soliton complexes becomes possible in the highly dispersive sine-Gordon system. Beginning the study of this phenomenon, first of all we were convinced that the exact solution (16) with the velocity (17) is absolutely stable and the soliton complex propagates without any radiation. If one chooses the initial profile in the same form but the velocity of the complex is rather small (e.g., $V_{in} = 0{,}3$) then the complex is dissociated in the explosive manner as indicated in Fig.5. Note that the complex center has been placed initially at the position $x = 20$. However the strong repulsion of kinks has caused their mutual fast escape in opposite directions with the velocities $V_1 \approx -0{,}25$ and $V_2 \approx 0{,}62$, respectively. We have also investigated a possibility of realization of the excited soliton complex states as it has been predicted in [13]. Indeed we have found such a situation in the weak dispersive case for $\beta = 1/12$ and $V_{in} = 0{,}5$. It is remarkable that this excited soliton complex survives (Fig.6) and appears to be stable.

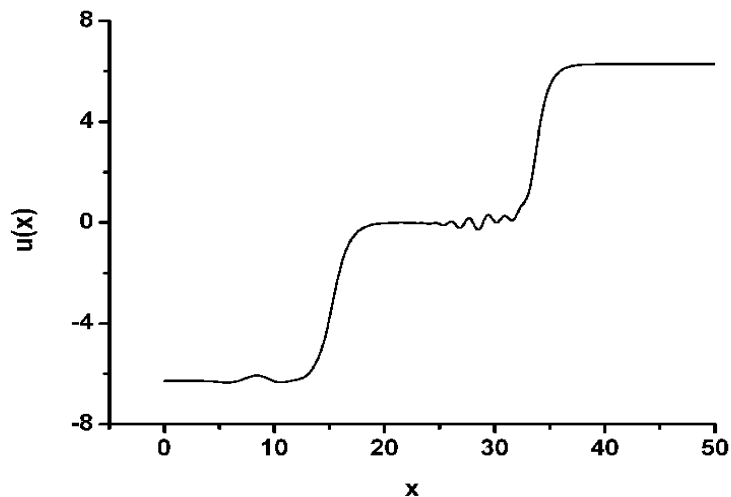

**Fig.5.** *The soliton complex dissociation in the case of a small initial velocity ( $V_{in} = 0,3$ ) in the dispersive sine-Gordon equation.*

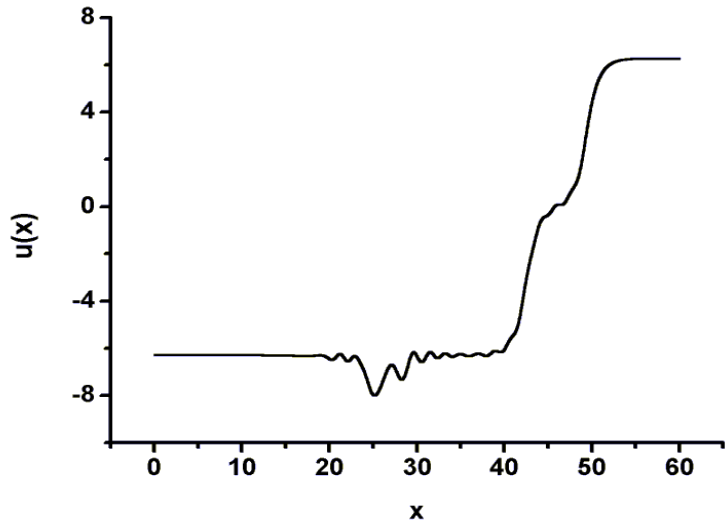

**Fig.6.** *The stable excited soliton complex state and the breather mode in the dispersive sine-Gordon equation.*

## Conclusions

We have investigated the internal kink dynamics and the soliton interaction in the dispersive sine-Gordon system, and the kink evolution in the dispersive $\varphi^4$-model. For the first time we have found analytically all the spectra of linear excitations of static kinks in the models and have studied numerically a large variety of the dynamical effects caused by the influence of the dispersion, which are inherent in nonlinear dynamics of topological excitations in lattices and continuous systems with non-local interactions. We have shown that the proposed models adequately describe main features of one-dimensional dispersive microstructured systems and macroscopic lattice-based media. They can be considered as very attractive objects for a study of novel universal effects in the nonlinear dispersive media. It should be noted that a lot of interesting phenomena has appeared to be beyond the scope of this study. In particular, there is a problem of description of the breather properties in the dispersive systems with a finite band spectrum. Now this work is in progress.